\documentclass[USenglish,oneside,twocolumn]{article}

\usepackage[utf8]{inputenc}
\usepackage[big]{dgruyter_NEW}
\DOI{foobar}
\usepackage[table]{xcolor}
\newcolumntype{R}[1]{>{\raggedleft\arraybackslash }b{#1}}
\newcolumntype{L}[1]{>{\raggedright\arraybackslash }b{#1}}

\usepackage{subcaption}
\cclogo{\includegraphics{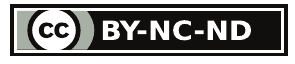}}
\begin{document}

  \author*[1]{Florentin Rochet}
  \author[2]{Olivier Pereira}

  \affil[1]{Université catholique de Louvain - ICTEAM - Crypto Group, E-mail: florentin.rochet@uclouvain.be}

  \affil[2]{Université catholique de Louvain - ICTEAM - Crypto Group, E-mail: olivier.pereira@uclouvain.be}

  \title{\huge Waterfilling: Balancing the Tor network with maximum diversity}

  \runningtitle{Waterfilling: Balancing the Tor network with maximum diversity}


  \begin{abstract} {We present the \emph{Waterfilling} circuit selection method, which we designed in order to mitigate the risks of a successful end-to-end traffic correlation attack. Waterfilling proceeds by  balancing the Tor network load as evenly as possible on endpoints of
      user paths. 
     We simulate the use of Waterfilling thanks to the TorPS and Shadow tools. Applying several security metrics, we show that the adoption of Waterfilling considerably increases the number of nodes that an adversary needs to control in order to be able to mount a successful attack, while somewhat decreasing the minimum amount of bandwidth required to do so. Moreover, we evaluate Waterfilling in Shadow and show that it does not impact significantly the performance of the network.
Furthermore, Waterfilling  reduces the benefits that an attacker could obtain by hacking into a top bandwidth Tor relay, hence limiting the risks raised by such relays. 
     Waterfilling does not require any major change in Tor, and can
     co-exist with the current circuit selection algorithm.  }
\end{abstract}
  \keywords{Tor, Path selection algorithm, anonymity, traffic correlation}

  \journalname{Proceedings on Privacy Enhancing Technologies}
\DOI{Editor to enter DOI}
  \startpage{1}
  \received{..}
  \revised{..}
  \accepted{..}

  \journalyear{2015}
  \journalvolume{2015}
  \journalissue{2}

\maketitle

\section{Introduction}
Tor is an implementation of an Onion Routing protocol designed to
provide anonymity over the Internet to TCP-based applications. Tor can
be seen as a distributed overlay network run by volunteer-operated
nodes where the users have an interest in remaining anonymous when
surfing on the web. Over the last few years, the Tor network has grown
from about 2000 to roughly 7000 relays~\cite{tormetrics}.  It provides
anonymity by bouncing the traffic through the network using a path of
at least 3 relays among the pool of nodes and rotates the path every
10 minutes~\cite{Tor-draft}. 

However, anonymity is not a guaranteed
property. Tor is known by design to be unable to preserve anonymity in
a situation where an adversary observes the traffic entering and
leaving the anonymity network. It is indeed possible for an adversary
to retrieve the identity of a Tor user and his destination by matching
the traffic at both endpoints of the anonymity network. This attack is
called end-to-end traffic
correlation and is considered in the literature to be a serious threat against anonymity networks. Many works were conducted to demystify the consequences of this threat over the Tor network. Global passive adversaries have been studied with different correlation functions to match streams entering and exiting the anonymity network. From simple packet counting~\cite{back01, SS03} to timing analysis~\cite{timing-fc2004, pet05-bissias}, the anonymity provided by the Tor network has showed to be circumvented~\cite{o2009improving}. However, such adversaries are too powerful and are not part of the Tor original threat model~\cite{Tor-draft}. More realistic passive adversaries have been studied, where either Autonomous Systems (ASes) or internet exchange points (IXPs) are compromised~\cite{feamster:wpes2004, murdoch-pet2007, DBLP:conf/ccs/EdmanS09}. Experimental results have shown a probability of 20\% to encounter the same AS at both ends of a path inside the Tor network. Johnson \textit{et al.}~\cite{ccs2013-usersrouted} have also shown a probability of 95\% to be deanonymized in three months by a single IXP for Tor users located in common places.

Monitoring parts of the Internet is not the only way to perform traffic correlation: since the Tor network is designed to be a volunteer-based network, an attacker can  also deploy nodes and use them to perform traffic correlation, waiting for Tor users to pick these  nodes at the ends of their path. Johnson \textit{et al.}~\cite{ccs2013-usersrouted} studied the likelihood of such a comprised path when realistic corrupted nodes are injected into the network. Results shows a probability of 80\% to encounter such a compromised path within a 6-month period between 2012 and 2013 for a 100MiB/s relay adversary. 

These results show the importance of fighting traffic correlation. One
approach is to design counter-measures tailored to the traffic
characteristics used to match streams~\cite{timing-fc2004,
  ShWa-Timing06}. While this approach has the potential of completely
suppressing the feasibility of one specific traffic correlation
attack, it also leaves completely open the possibility to use a
different correlation measurement method. A second approach, which we
explore in this paper, consists in designing the network in such a way
that it minimizes the probability that an adversary could monitor both
ends of a path through the Tor network. This second approach cannot
fully preclude the feasibility of a correlation attack, but it is more
robust than the first one in the sense that it is independent of the
correlation detection method that is used. The two approaches can of
course be combined and can, in some specific circumstances, interfere
with each other~\cite{ccs2011-stealthy}.

Modifying the path selection algorithm in order to reduce the feasibility of a traffic correlation attack is most appealing if the overall performance of the network stays unchanged. Currently, Tor uses a selection process called Adjusted Bandwidth Weight Random Selection (ABWRS) that combines bandwidth-weights and nodes perceived bandwidth to apply the weighted choice. The perceived bandwidth allows Tor clients to bias toward relays with more available resources while the bandwidth-weights allow to balance the load from one position to another (e.g., if there is too much bandwidth for the entry position, a fraction of each node bandwidth is moved to the middle position). 



\paragraph*{Our contributions}

In this paper, we explore a different way to achieve the balance,
providing a more uniform relay selection mechanism to Tor users.
Waterfilling replaces the single fraction of the bandwidth attributed to all relays for each (available) position to a fraction defined per relay. While not impacting the total bandwidth of the network, these individual fractions  make sure that low-bandwidth relays devote most of their capacity to traffic at the endpoints of circuits, making it possible for high-bandwidth relays to devote a larger part of their capacity to the middle-point of circuits. As a result, top relays become a considerably lower threat to anonymity compared to the current Tor network state.

More precisely, in this paper: 
\begin{itemize}
\item We suggest a modification of the current Adjusted Bandwidth
  Weight Random Selection (ABWRS) called Waterfilling Adjusted
  Bandwidth Weight Random Selection (WFABWRS) or Waterfilling for
  short. Waterfilling keeps the Tor network balanced and provides
  either more diversity in both ends or in one end of the Tor network,
  depending on the network load case.
\item In order to evaluate the impact of Waterfilling, we use known
  metrics and propose a new one, based on guessing entropy~\cite{Massey94guessingand}. Our metric indicates the expectation on the number of
  nodes to be compromised before being able to mount a successful correlation attack, and
  we believe that it provides useful information on the security of
  the Tor network at a given time.
\item We provide a concrete analysis of the benefits of Waterfilling
  based on simulations executed from consensuses available for 5
  months during 2015.  Our analysis shows that an attacker would be
  required to control on average 150 extra nodes in order to run a
  successful end-to-end correlation attack on a targeted circuit, and
  that he would need to control 35 nodes in order to obtain the same
  benefits as if he had compromised the top guard node during the first half of
  2015.

\item We modified TorPS~\cite{torps} in order to implement
  Waterfilling and, on our way, we identified and fixed an
  issue, 
  which caused the results in previous publications to
  overestimate the success probability of attacks. Our pull request has been merged in the original github project.
\item We developed a prototype of Waterfilling in the Tor 0.2.6.7 base code and we performed concrete performance evaluations in the Shadow simulator, comparing Waterfilling with Tor's ABWRS.
\end{itemize}


\paragraph*{Roadmap}
We start by reviewing the related works regarding path selection algorithms and anonymity metrics in Section \ref{sec:related_work} and provide background information in Section \ref{sec:background}. Then, we dive into our subject in Section \ref{sec:waterfilling} explaining what is Waterfilling and how we compute our new bandwidth-weights. We then describe in Section \ref{sec:models} our threat model and explain which metrics will be used and why they are suitable. Our Section \ref{sec:security} gives the security analysis of Waterfilling versus unmodified Tor using our metrics and an empirical evaluation against relay adversaries. Our Section \ref{sec:performance} gives the performance analysis of Waterfilling in terms of expected circuits latency and time to download a web page or a large file. 
 We discuss why Waterfilling should be easy to deploy in Section \ref{sec:deployment}, and we discuss current limitations and open questions in Section~\ref{sec:limitations}. We conclude in Section \ref{sec:conclusion}.

\section{Related Work}
\label{sec:related_work}
\paragraph*{Path selection algorithms}
%

In the original Tor design, the basis of the path selection algorithm was to select uniformly at random the nodes needed to build a circuit. The reason to do this was to meet the theoretical security level of Onion Routing against relay adversaries~\cite{onion-routing:pet2000}. However this strategy does not provide a good network performance for Tor users in average, because relays offer very different bandwidths. Therefore, the selection moved to an adjusted bandwidth weight random selection choice, making the selection probability of a node proportional to the bandwidth it offers (see details in Section \ref{sec:path_selection_details}). Several other proposals of path selection algorithms have been made during the past few years.


Snader and Borisov studied a tune-up to balance between the anonymity and performance properties~\cite{snader08}. LASTor from Akhoondi \textit{et al.}~\cite{oakland2012-lastor} suggests a path selection algorithm that reduces both the latency and the risk to encounter a circuit where both entry and exit are in the same Autonomous System (AS). Unfortunately, LASTor does not result in a balanced network and suffers from the same throughput issue as uniform selection. Hence, it cannot be deployed as it is suggested~\cite{ndss13-relay-selection}.

In a complementary way, some methodologies have been developed to improve performance regarding latency on the top of a path selection algorithm. 
In this line of work, the ``normal'' path selection algorithm (or another one) is used as a first step, but a second selection step is proposed in order to filter among paths obtained in the first step. 
Wang \textit{et al.} ~\cite{congestion-tor12} developed a congestion-aware selection scheme that models the congestion of a circuit as a node-based approach instead of a link-based approach. In this case, the filtering is on the top of the normal Tor adjusted bandwidth weighted random selection. An interesting fact is that the congestion computed for the nodes is not correlated with their bandwidth, hence such a scheme does not unbalance the network. Sherr \textit{et al.} proposed a latency-aware link-based path selection algorithm called Coordinate on the top of Snader and Borisov's scheme~\cite{DBLP:conf/pet/SherrBL09}. More recently, Wacek \textit{et al.} suggested hybridizing the ``normal'' path selection algorithm with Coordinate and evaluated its performance~\cite{ndss13-relay-selection}.

Hybridizing path selections or looking to improve an existing idea have been welcomed and even integrated in Tor in the past few years. In this line of work, we have the Guard flag that is a response to protect against the Predecessor Attack introduced by Wright \textit{et al.}~\cite{Wright:2004,torblog}. Researchers evaluated also the well-funded values of some policies. Elahi \textit{et al.}~\cite{wpes12-cogs} simulated various guard related parameters to assert which approach might be more interesting for Tor users, and Backes \textit{et al.}~\cite{backes2015your} also explored various path selection strategies.  


\paragraph*{DistribuTor}


The closest proposal of our work is the suggestion by Backes
\textit{et al.} at CCS'14~\cite{ccs2014-mators}.  They present a path
selection called DistribuTor that redistributes the bandwidths based
on computation performed on the client side (and not by the Tor
authorities). In DistribuTor, a bandwidth upper bound is chosen by the
client and the choice of an exit node is based on both the bandwidth
available on the exit node and on that upper bound: the probability of
selecting an exit node is proportional to its bandwidth, except that
this bandwidth is considered to be trimmed to the chosen upper bound,
effectively bounding the probability of selecting high bandwidth
nodes. 

Our work differs from that one in several aspects: 
\begin{itemize}
\item We focus on entry point, which is more important
  for protection against correlation attacks. 
\item We push the definition of the probability selection of the
  various nodes back to the directory authorities, which helps
  maintaining a balanced network.
\item We perform detailed security and performance analysis of our path selection algorithm. 
\end{itemize}
\paragraph*{Anonymity models and metrics}

The general consensus from the literature regarding modelling anonymity comes from Pfitzmann and Hansen~\cite{terminology}  and holds in a set of definitions. First of all, anonymity is defined as the state of not being identifiable within a set of subjects called the anonymity set. The anonymity set stands for the set of all probable subjects. These subjects are linked to anonymous actions which remain anonymous if an adversary cannot distinguish the subjects on which they are executed. The definitions of the subjects and their related anonymous actions are context dependent and defined regarding the anonymity system studied. Surveys over anonymous systems were conducted in~\cite{edman2009survey, systems-anon-communication}, including mix networks, Dining Cryptographers networks and Onion Routing.

To quantify anonymity, we use probability distribution about the set of subjects linked to the anonymous actions we study. From these probability distributions, we extract information using the notion of entropy. Entropy provides a quantification of the uncertainty involved in predicting on which subject is linked our anonymous action. Depending on which information we extract, we infer different meaning of the result.

Serjantov and  Danezis~\cite{Serj02} and Diaz \textit{et al.}~\cite{Diaz02} are the first to suggest the idea of using information theory metrics such as Shannon's entropy~\cite{shannon2001mathematical} to infer the number of bits of additional information that the attacker needs in order to definitely identify the subject. Moreover, Diaz \textit{et al.} suggested using a normalized version of Shannon's entropy as the degree of anonymity for the system. A degree 0 means that the anonymity system does not provide anonymity at all. A degree 1 means that the uncertainty is maximal. Snader and Borisov~\cite{snader08} suggested using the Gini coefficient as a measure of equality for the subjects. A Gini coefficient of 0 means that we have perfect equality between the subjects. we cannot distinguish one from the other regarding our anonymous action. A coefficient of 1 means a perfect inequality. 
Those metrics have been extensively used in the literature by the Tor community. 

However, even if entropy is widely used, there are some drawbacks to it. For instance, Syverson pointed that Shannon entropy, as an average, does not necessarily capture worst case situations~\cite{Syverson_why}. If we take the following example where we have a set of 1025 potential senders of a message with 1024 of them having an equal probability of 1/2048 to be the origin of the message, and a single sender with probability of 1/2. We end up having a degree of anonymity of 0.6 for this distribution which is quite high regarding the fact that one sender has a strong probability to be the source. 

In the Tor community, some researchers started to advocate the use of an empirical anonymity measure based on a well-defined adversary. Hamel \textit{et al.} suggested measuring the probability of path compromise under an adversary with fixed bandwidth capability~\cite{hamel2011misentropists}. Johnson \textit{et al.} built a Tor Path Simulator to mimic the Tor path selection over time and infer statistical confidence about first path compromise under a fixed relay adversary~\cite{ccs2013-usersrouted}. We also use such metrics in our anonymity evaluation.\\
%
%


%

\section{Background}
\label{sec:background}
The Tor network is composed of different types of nodes, \textit{onion proxies} (Tor client), \textit{onion routers} (relays), \textit{directory servers}, \textit{bridges} and \textit{hidden servers}.  Directory servers are responsible for setting the network up by publishing a consensus document every hour that assigns a selection weight for each relay role. Those weights are used to balance the network between the three node positions respectively entry, middle and exit nodes. A relay might have different roles and could act as an exit but also as an entry and a middle node. 
Thus, the weights allow the resources of a relay to be proportionally distributed among the different roles that it handles.
The nodes are divided among the positions according to some status flags assigned by the \textit{directory servers}. We have the \textit{Guard} flag that allows a relay to be picked out by the Tor client as its entry node in the Tor network. Guard nodes must fulfill performance and stability constraints, and we currently have a pool of roughly 1600 Guards among all relays. An \textit{Exit} flag is also assigned to nodes that accept exit policies for range of IP addresses and ports. Exit nodes are picked out by the Tor client to be the node responsible to connect to the requested service. The destination IP and port must match the node exit policy, hence the pool of available exit nodes depends on the service that the user wants to access. All remaining relays are middle nodes.
 
\subsection{Path selection in details}
\label{sec:path_selection_details}
The description above showed that we have two entities involved in the path selection: directory servers and Tor clients.
\paragraph*{Directory servers}

 The Tor Project provides documentation about Tor specifications~\cite{dirspec} and explains the responsibilities of directory servers. Among them, the weight computation is our subject of interest. Weights are used to balance the network among the positions, which arise from the solution of a simple system of equations related to the equality of the bandwidths from entry, middle and exit positions. 
 The system of equations from dir-spec.txt is:
\begin{small}
\begin{align}
	Wgg.G+Wgd.D & = &  M+Wmd.D+Wme.E+Wmg.G \label{eq:1}\\
 	Wgg.G+Wgd.D & = & Wee.E + Wed.D \label{eq:2}\\
 	D & = & Wed.D+Wmd.D+Wgd.D \\
 	G & = & Wmg.G + Wgg.G\\
 	E & = & Wme.E+Wee.E
\end{align}
\end{small}
With:
~\\
\begin{itemize}
\item $G$ being the total bandwidth for Guard-flagged nodes
\item $M$ being the total bandwidth for non-flagged nodes
\item $E$ being the total bandwidth for Exit-flagged nodes
\item $D$ being the total bandwidth for Guard+Exit-flagged nodes
\item $Wgd$ being the weight for choosing a Guard+Exit for the guard position
\item $Wmd$ being the weight for choosing a Guard+Exit for the middle position
\item $Wed$ being the weight for choosing a Guard+Exit for the exit position
\item $Wme$ being the weight for choosing an Exit for the middle position
\item $Wmg$ being the weight for choosing a Guard for the middle position
\item $Wgg$ being the weight for choosing a Guard for the entry position
\item $Wee$ being the weight for choosing an Exit for the exit position
\end{itemize}

This system constrains a bandwidth equality between entry and middle
nodes in equation (1). The same constraint is applied between middle
nodes and exit nodes in equation (2). Equations (3), (4) and (5)
ensure that the weights are consistent. Of course, these equations
cannot always be satisfied, when a resource (e.g., exit nodes) is
scarce in the network. In such cases, some of these equations become
inequalities, and the bandwidth is allocated on a case-by-case basis:
the Tor specification divides all possibilities into 12 cases, and
provides constraints for each of them.


It is worth noticing that this system of equations achieves a balanced network condition regardless of the bandwidth of each node independently. It only cares about the sum of each pool ($G$, $M$, $E$, $D$). It results that if we have to balance bandwidth from one position to another, the same fraction of bandwidth is transferred for each node. 

\paragraph*{Tor clients}
Once the weights have been computed, voted and published in a network status document, the Tor client uses them to assign selection probability to each relay. Each Tor client biases its selection according to the weights received for each position and the consensus weight \footnote{The consensus weight is called Bandwidth in the network status document} of the relays.
We have:
$$ ClientWeight_{(relay_i, position~p)} = ConsensusWeight_i * W_{pf} $$
With $W_{pf}$ the weight computed by the directory server, depending
on the desired position and the flags of $relay_i$. Then, the Tor
client makes a weighted random choice among relays when building
circuits, using all computed $ClientWeights$. Consequently, if each
Tor user applies this strategy, the Tor network end up having the same
bandwidth consumed by relays for entry, middle and exit
positions\footnote{Unless we fall into a scenario where the resource
  of a position is too scarce}. Also, the ClientWeight of a relay
depends directly on its consensus weight which is a value based on the
perceived bandwidth of the relays measured by the directory
servers. We end up having a selection probability of a relay for a
specific position that is directly proportional to its perceived
bandwidth:

\begin{align}
\label{eq:6}
Pr[relay_i, position~p] = \frac{ConsensusWeight_i * W_{pf}}{\sum_j ConsensusWeight_j * W_{pf}}
\end{align}
where the sum at the denominator ranges over all nodes  and $W_{pf}$ is defined to be 0 for nodes that cannot be placed in position $p$. 


There is however a little extra complexity in the selection process, as some constraints
are placed on the structure of a circuit: 

\begin{itemize}
\item The exit node must have a policy accepting connections to the desired IP address and port; 
	\item The same relay can not be chosen twice for the same circuit;
	\item Two relays in the same family can not be chosen for the same circuit;
	\item At most one relay selected in a given /16 subnet for the same circuit.
\end{itemize}

We will not touch these constraints here. 

%

\section{Waterfilling Bandwidth-Weights}
\label{sec:waterfilling}
We suggest changing the bandwidth weights computation method in order to maximize the diversity selection in guard and exit position. Before diving into the equations and generalizing the concept, we show in the first subsection the intuition behind our Waterfilling strategy with an example.
	\subsection{Looking at the big picture}
\begin{figure}
	\centering
	\includegraphics[scale=0.45]{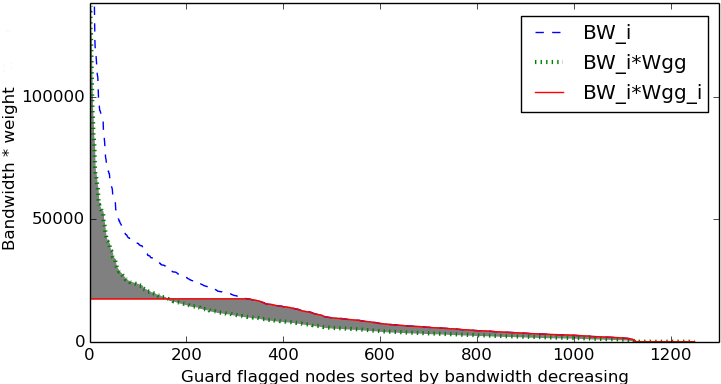}
	\caption{Waterfilling on guard nodes for the 10th consensus from 25th May 2015}
	\label{wf_example}
\end{figure}

 Figure~\ref{wf_example} shows all guard-flagged nodes sorted by decreasing consensus weight (bandwidth). The dashed blue line shows the total capacity of each guard-flagged node received from the network status document. The dotted green line shows the capacity of each guard-flagged node dedicated to the entry position, which is decided by the $Wgg$ value from the network status document. The capacity between those two lines is what is transferred to the middle position to verify the equations previously introduced in Section~\ref{sec:path_selection_details}. It is a shift of resource from guards to unflagged nodes to obtain the equivalence of total bandwidth consumption from each position.
 
The plain red line is the result of our Waterfilling scheme. It allows transferring the same amount of capacity to the middle position by moving from a global $Wgg$ to a per-node $Wgg_i$. Everything above the horizontal red segment (the water level) is transferred to the middle position. That is, we fill the smaller guard-flagged nodes until a level where all above area enclosed between the dashed blue line and the horizontal part of the plain red line is equal to the area enclosed between the dashed blue and dotted green lines. We allocate resources differently but we shift the same amount of capacity as with classic bandwidth-weights, thus leaving the network capacity untouched. However, by using lower-bandwidth guards fully in their guard role, and by capping the use of higher-bandwidth nodes for guard traffic, we obtain a much more uniform probability distribution for guard node selection, which will render correlation attacks more challenging to mount. 

The network load case for this particular example corresponds to a
situation in which the bandwidth in the exit position is scarce and
the total bandwidth of guard position is greater than the total
bandwidth of middle position. This situation corresponds to the third
case, subcase $a$ in the Tor specification, and we write it
\textit{3aE=SG>M}. It is the most representative, and has been
observed $97.7 \%$ of the time during the first five months of
2015. For the remaining percentage, we are in a single network case, for which we propose a similar Waterfilling strategy. 

\subsection{Computing Waterfilling bandwidth weights}
\label{sec:waterfilling-weights}

Following our example above, we first explain how to compute the Waterfilling bandwidth weights through the network load case \textit{3aE=SG>M}. For this network load case, the Tor specification indicates to compute the weights from the following equations: \\
\begin{align}
Wee = Wed = 1; \label{eq:7} \\
Wmd = Wgd = Wme = 0; \label{eq:8} \\
Wgg = (G+M)/(2*G) \label{eq:9} \\
Wmg = 1 - Wgg; \label{eq:10}
\end{align}


We can apply Waterfilling to $Wgg$ when $Wgg$ is neither $0$ nor $1$. Generally speaking, with another network load case, we may also end up with $Wee$ or $Wed$ or $Wgd$ not equal to $0$ or $1$. If this is the case, the following introduced constraints can be symmetrically applied to $Wee$ and with a small modification to $Wed$ and $Wgd$.

Now, based on these values, we compute the individual weights $Wgg_i$
of each guard node, from which we can also derive the index $N$ of the
\emph{pivot} node, which the last guard to be ``above the water
level'', or the last guard that is going to be used both as guard and
middle node. It is the point where the dashed blue line and the
horizontal part of the plain red line meet in Figure~\ref{wf_example}.

Suppose there are $K$ nodes with the guard flag, and let $BW_i$ be the
bandwidth of the $i$-th guard node with the highest bandwidth. Then
%
the new constraints are:
\begin{align}
	Wgg_i*BW_i = Wgg_{i+1}*BW_{i+1} ~\forall i \in (1, N) \label{eq:11}
\end{align}
\begin{align}
	Wgg_i = 1 ~\forall i \in (N+1, K) \label{eq:12}\\
 	0 \leq Wgg_i \leq 1 ~\forall i \in (1, K) \label{eq:13}\\
	\sum_{i=1}^{K} Wgg_i BW_i = Wgg*G \label{eq:14}
\end{align}


Equation~\ref{eq:11} expresses that all nodes before the pivot must
devote the same amount of bandwidth to the guard
position. Equation~\ref{eq:12} expresses that all nodes after the pivot
position will fully play their role as guards.  Equation~\ref{eq:13}
guarantees that no node will be required to offer more bandwidth than
available, and Equation~\ref{eq:14} guarantees that the total amount
of bandwidth available for the guard position remains unchanged
compared to the original Tor strategy, based on the unique $Wgg$. From
a visual point of view, this equation guarantees that both grey areas
have equal surface in Figure~\ref{wf_example}.

Solving with these constraints gives the weight $Wgg_i$ of each node, from which we compute $Wmg_i=1-Wgg_i$ for each guard-flagged node.

\subsection{Going further with Waterfilling}
\label{sec:further_wf}
\begin{figure}
	\centering
	\includegraphics[scale=0.45]{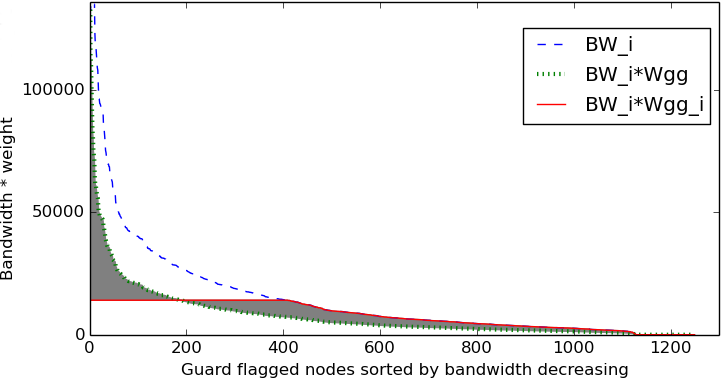}
	\caption{Better Waterfilling obtained from a recalculation of the bandwidth-weights. Waterfilling is applied on guard nodes for the 10th consensus from 25th May 2015}
	\label{wf_example_2}
\end{figure}

Waterfilling balances the network on the top of classical bandwidth-weights. We may wonder if those bandwidth-weights are always computed in a suitable way for Waterfilling. In some scenarios, the equations \ref{eq:1} and \ref{eq:2} from Section~\ref{sec:background} cannot be verified together: when either $(G+D) < T/3$ or $(E+D) < T/3$.\footnote{T being the total bandwidth: T=G+M+E+D} In the previous section, we discuss a consensus where we have the second scenario: $(E+D) < T/3$. In this scenario, we cannot achieve a balance between the three positions. The way Tor specifications suggests computing the weights for this particular scenarios (Equations \ref{eq:7}, \ref{eq:8}, \ref{eq:9}, \ref{eq:10}) shows that they verify the equation \ref{eq:1} but release equation \ref{eq:2} to an inequality. We have:
\begin{small}
\begin{align}
	Wgg.G+Wgd.D & = &  M+Wmd.D+Wme.E+Wmg.G \\
 	Wgg.G+Wgd.D & > & Wee.E + Wed.D 
\end{align}
\end{small}
Therefore, they equalize the bandwidth between guards and middles. We may achieve a better balance for Waterfilling: equalizing the bandwidth between exits and guards and pushing what remains to the middle position. The equations become:
\begin{small}
\begin{align}
	Wgg.G+Wgd.D & < &  M+Wmd.D+Wme.E+Wmg.G \\
 	Wgg.G+Wgd.D & = & Wee.E + Wed.D 
\end{align}
\end{small}
Achieving the balance this way reduces the $Wgg$ value compared to the first approach. It becomes:
\begin{align}
Wgg = (E+D)/G 
\label{eq19}
\end{align}

We can compare Figure~\ref{wf_example} and Figure~\ref{wf_example_2} when Waterfilling is applied on these two approaches. On Figure~\ref{wf_example}, the pivot is around the 340th node while on Figure~\ref{wf_example_2}, it is around the 400th node. Consequently, the water level is smaller than with the approach from Tor specifications. Hence, the probability selection on guard nodes are closer to the uniform distribution. We may also wonder if this approach reduces the performance of the network. Intuitively, we have now two positions where congestion might occur instead of one. We answer this concern in our performance analysis explained in Section \ref{sec:performance}.

\section{Security Models and Metrics}
\label{sec:models}

\subsection{Threat Model}
	
	Since its initial design, Tor has been known to lack  efficient protection against end-to-end traffic correlation. Solving this problem might be impossible or at least extremely difficult for any low-latency anonymity network. Existing techniques such as padding~\cite{ShWa-Timing06} have shown to be too costly to be deployable while other techniques such as delaying cells or defensive dropping~\cite{timing-fc2004} have shown to be ineffective in practice. One way to prevent end-to-end traffic correlation is to accept it and to strive to minimize its impact while guaranteeing a high quality of service. 
	
	In this threat model, an adversary controls a bunch of nodes in order to perform end-to-end traffic correlation when a Tor user is passing through one of its guards and exits nodes. Following Murdoch \textit{et al.}~\cite{murdoch-pet2008}, a realistic view of a relay adversary considers both  IP addresses and bandwidth to contribute to a cost that the attacker tries to minimize. We consider two variants of this model:
	\begin{itemize}
		\item Budget relay adversary: An adversary deploying its own relays into the network. The budget is fulfilled by the cost of the bandwidth and the IP addresses. In order to avoid obvious detection, a relay adversary would take different /16 if he chooses to deploy several nodes, and distribute them into different data centers, hence increasing its costs proportionally to the number of nodes and total bandwidth. 
		\item Intruder relay adversary: An adversary hacking into existing Tor relays, servers or private computers. In this case, the bandwidth has little direct influence on the cost, since the attacker does not pay for it. The cost will instead be related to the number of machines that need to be corrupted, and to their level of protection. Here, a generic botnet is unlikely to offer a good strategy for an adversary, due to the regular downtime of the corrupted machines~\cite{stone2009your}, and more efforts will then be needed. At the other extreme of the spectrum, big existing relays are likely to be top targets for such an adversary, at least in the current Tor path selection algorithm: these relays already handle a major part of the traffic due to the current path selection algorithm. Besides, the visible effects of a corruption will be lower, since this corruption does not add a new relay traffic on the network. In all cases, due to the relatively high profile of the machines to be corrupted (they need to be able to obtain the guard flag), the number of machines to corrupt appears to be an important measure of the adversary effort.  
	\end{itemize}

        Based on these observations, we will evaluate the cost of a
        successful correlation attack both in terms of relays that
        need to be corrupted and total bandwidth that is required, the
        relative weight of these depending on the exact setting and
        willingness of the adversary to hack into other computers.

We make the assumption that a correlation is instantaneous and perfect. 
	Notice that network adversaries are not taken into account in this model.
	





\subsection{Metrics}

In order to evaluate the impact of our Waterfilling strategy, we use
three anonymity metrics.

Our first metric, used in related works, is the \textit{uniformity
  degree of circuit selection}, computed as Diaz's degree of
anonymity~\cite{Diaz02} over the probability distribution of selected
guard-exit pairs in the Tor network. This metric gives an indication
of how well the network is exploited at a given point in time, and is
normalized (that is, independent of the network size). 

Our second metric is based on the notion of \textit{guessing
  entropy}~\cite{Massey94guessingand}, and indicates the expected
number of nodes that an adversary should control in order to mount a
successful end-to-end traffic correlation attack. This metric is
related to the previous one in the sense that it also focuses on the
state of the network at a single point in time, but it provides a more
concrete information to the user.

The last metric we use is the \textit{probability distribution on time
  until first path compromise}, which is an empirical metrics from
Johnson \textit{et al.}~\cite{ccs2013-usersrouted}.  Contrary to the
other two, this metric adopts a dynamic perspective by considering,
across a certain amount of time, the success probability of an
adversary who controls a predefined number of nodes and offers a
predefined bandwidth on these nodes. 

\subsubsection{Uniformity degree of circuit selection}

The uniformity degree of circuit selection measures how
close to uniform is the selection process of a guard and an exit
node for a circuit. It shows a first interesting indication of the
resistance of the network to end-to-end correlation attacks.

This uniformity degree is computed by evaluating the probabilities
$p_{i,j}$, which indicate, for all $i$ and $j$, the probability of
picking the guard $i$ with exit $j$ in a circuit. Then the Shannon
entropy of this distribution is computed as:
$H(Y) = - \sum_{i=1}^{N}\sum_{j=1}^{K} p_{i,j}\log_2(p_{i,j})$.
 
This quantity is then normalized to the maximum entropy $H_M$ that this distribution could have, which is computed as $log_2(N*K)$, where $N$ and $K$ are the size of the set of guard and exit nodes respectively. 

So, the uniformity degree $d$ of a circuit selection process is computed as: 
$d = \frac{H(Y)}{H_M}.$

Note that it may not be desirable to obtain a uniformity degree of $1$, as it would not take into account any topological aspect of the circuit, for which policies are in place (e.g., we cannot have two same /16 addresses in the circuit).

The uniformity degree of circuit selection is interesting to compare the quality of a path selection algorithm on different states of the Tor network that do not necessarily contain the same number of nodes. 

We evaluate this metric by using the TorPS tool to run simulations of the original and Waterfilling based circuit selection processes, based on various consensuses from the first five months of 2015: these simulations offer estimations of the $p_{i,j}$ values, from which the uniformity degree can be evaluated as defined above. 

\subsubsection{Guessing entropy}

The uniformity degree provides an interesting indication on the
distribution of the circuit selection process, but does not provide a
measure that can easily be interpreted in terms of success probability
of an end-to-end correlation attack: Shannon entropy is an average
measure, not a worst-case one, and the normalization aspect makes it
possible to obtain a very high degree of uniformity even for a Tor
network that could contain 3 nodes. 

In order to address these questions, we propose using the notion of
\emph{guessing entropy}, as a measure of the number of nodes that an
adversary must expect to control or compromise before being
able to deanonymize a specific circuit.

The guessing entropy is computed from the same probabilities $p_{i,j}$
as above, by ranking the relays in decreasing order of their
contribution to the success probability of a successful end-to-end
correlation attack.

We then define $q_i$ as the marginal probability of a successful attack (that is, the increase of success probability of the attack resulting of the compromise of the $i$-th node), and evaluate the guessing entropy of the resulting distribution: 
$$g = \sum_{i=1}^{N+K}i.q_i$$

Of course, we always have $q_1 = 0$ because we cannot mount a correlation attack from a single compromised node, but we choose the first node in order to maximize the impact of the compromise of a second node. 
In order to compute the remaining elements of our vector $q$, we keep selecting nodes in a monotonic way such that $q_i$ is maximal with respect to the $i-1$ already selected nodes.
\begin{align}
Pr_{G=x} = max\left(\sum_{y \in q}Pr(G=x, E=y) \forall x \notin q\right)\\
Pr_{E=y} = max\left(\sum_{x \in q}Pr(G=x, E=y) \forall y \notin q\right)\\
q_i = max(Pr_{G=x}, Pr_{E=y})
\end{align}
An example is available in Appendix \ref{appendix:guessing}.  

This selection process corresponds to a monotonic strategy in which
the adversary compromises relays one after the other, looks for
the best choice based on the relays that he already compromised, and
does so until his attack succeeds. 

This strategy is definitely more effective than simply looking for
individual nodes from the two sets and constructing the product of the
distributions rather than our joint distribution, as shown by Johnson
et al.~\cite{johnson2009more}. A more effective attacker strategy
would be to select sets of relays to be jointly compromised instead of
picking them one by one, but such a strategy would also be
considerably more difficult to translate into a simple metric due to
combinatorial explosion.

\subsubsection{Time until first compromise}

This measure gives an estimate of the evolution over time of the probability until a first path compromise happens, for an adversary controlling a specific set of nodes.

To evaluate this measure, we repeatedly simulated the circuit
selection process of a client during a period of 5 months using TorPS,
and used these simulations to estimate the probability of  building a
compromised circuit over time. We applied this strategy with the
current Tor selection scheme and with our modified Tor Waterfilling
selection scheme, for comparison. More details about our relay
adversary are given in Section \ref{sec:security}.

This measure has the advantage of giving an interesting insight
regarding a concrete threat, as far as TorPS correctly mimics Tor
circuit selections. And, as a complement information to the metrics
discussed above, this one integrates a dynamic aspect of circuit
compromise over time, and not just at a given point in time, hence
taking into account path rotation and relay instabilities.

However, it is also a very specific measure, that is essentially valid
for a particular choice of relay adversary and the time period that is
chosen. 

\section{Security Analysis}
\label{sec:security}
We now evaluate our Waterfilling scheme by comparison with the current Tor path selection process, based on the three metrics discussed in the previous section. 

\subsection{Methodology}

To evaluate the security of the Tor network with respect to our threat
model, we need to compute the probability distribution of node
selection. There are multiple ways to do it. A first one would have
been to use the equations from Sections~\ref{sec:path_selection_details} 
and derive probabilities accordingly. However, the resulting distribution would not
account for the extra path construction policies (e.g., no pair
of nodes from the same /16 range on a single circuit) or for
realistic user behaviors (e.g., it does not take into account exit
policies that can also shape path selection). 

In order to obtain more realistic results, we used the TorPS tool
with the objective to evaluate the probability distribution considering the
typical user model from~\cite{ccs2013-usersrouted}. This user model
has been designed to mimic average Tor use with simulated connections
to Gmail / Google Chat, Google Calendar/Docs, Facebook and web search.

While working with TorPS, we noticed and fixed an issue where the
addition of relay adversary bandwidth was not considered in the
computation of the bandwidth-weights. Thus, adding nodes in the
simulated network made TorPS inaccuratly reflect Tor's path
selection: the higher the injected bandwidth, the further the TorPS
results diverged from the reality. As a result, our simulation results should not be directly compared to those obtained by Johnson \textit{et al.}~\cite{ccs2013-usersrouted}.

We studied the anonymity of the Tor network over the first 5 months of
the year 2015 \footnote{We wanted to update with first semester of 2016 but the Stem library used by TorPS is outputting many errors when reading descriptors. We got then many missing descriptors.}. Inside this range, we pick 20 given moments for each
type of network load/resource scarcity appearing in this period. From
these simulations we evaluate our various metrics. 

We also compare our Waterfilling scheme and Tor's ABWRS against a relay adversary. For both schemes, we compute the time until the first compromise path and we discuss different relay adversary strategies.
\subsection{Analysis}
\label{sec:subanalysis}
\begin{figure*}
	\begin{minipage}[t]{0.45\linewidth}
        \centering
	    \includegraphics[scale=0.295]{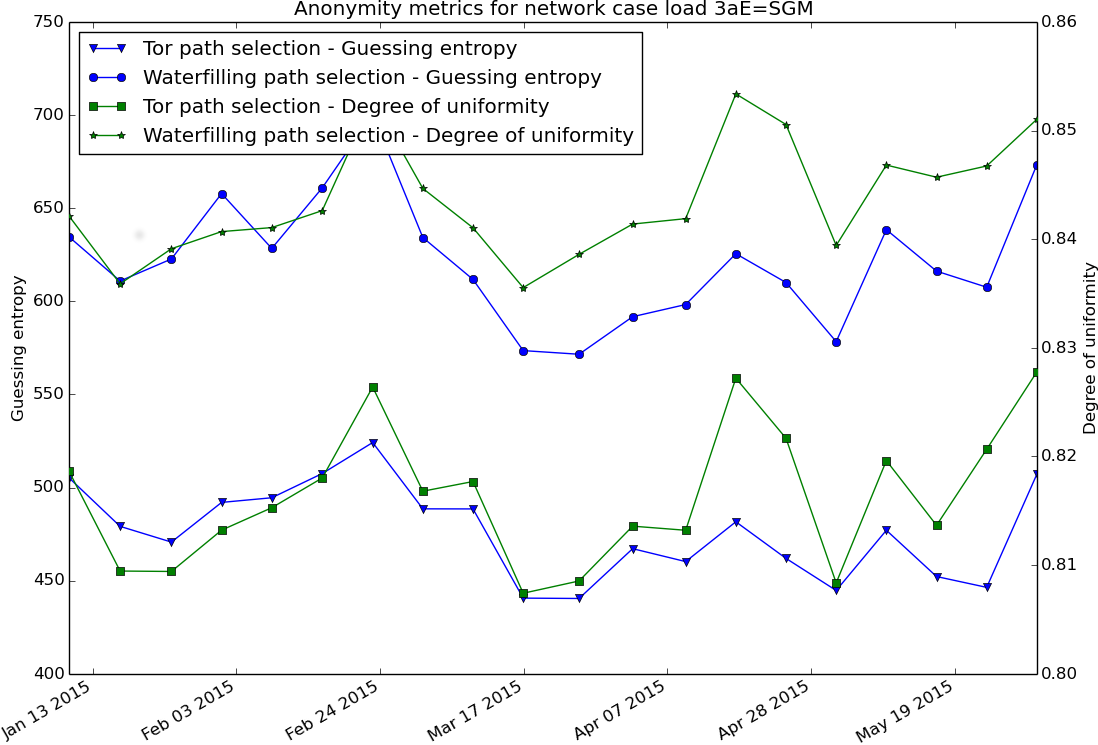}
	    \caption{Anonymity metrics over 20 given moments where network load case is 3a exit nodes are scarce and there is more bandwidth in Guard position than in middle position (3aE=SG>M) during 01-2015 to 05-2015. Waterfilling simulations use equations from Section \ref{sec:further_wf}}
        \label{fig:figure1}
	\end{minipage}
	\hspace{0.8cm}
	\begin{minipage}[t]{0.45\linewidth}
		\centering
		\includegraphics[scale=0.34]{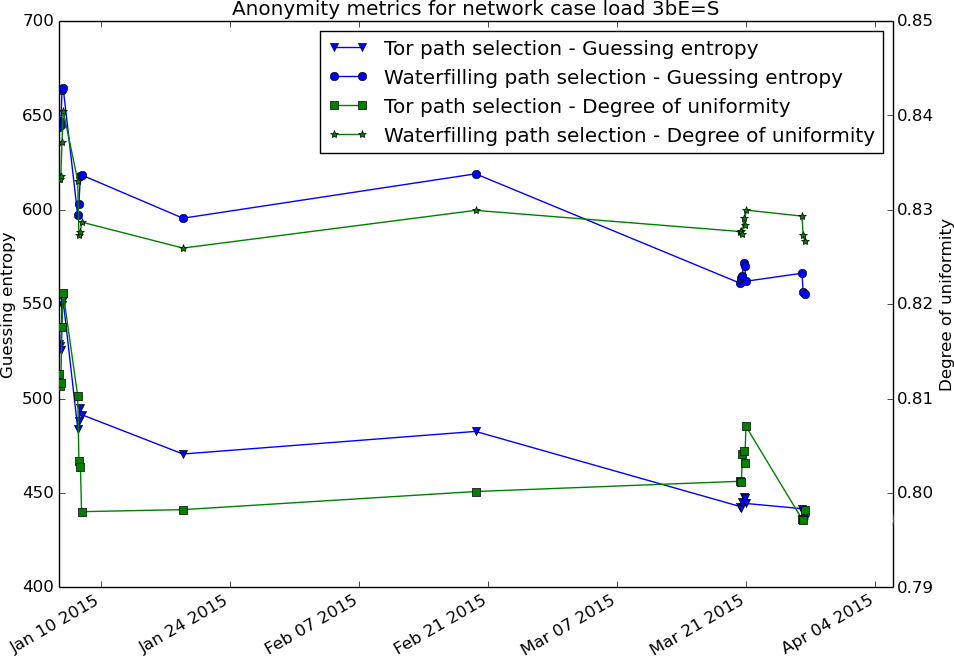}
	    \caption{Anonymity metrics over 20 given moments where network load case is 3b and exit nodes are scarce (3bE=S) during 01-2015 to 05-2015}
        \label{fig:figure2}
	\end{minipage}
\end{figure*}

During the first 5 months of 2015, only two of 12 possible network load cases have been observed: in both cases pure exit nodes were scarce ($E < T/3$); most of the time adding the nodes with the guard and exit flags was not enough to remove scarcity ($E+D < T/3$, case 3aE=SG>M) but, at a few moments, they were sufficient ($E+D \geq T/3$, case 3bE=S). Figures~\ref{fig:figure1} and~\ref{fig:figure2} show how the uniformity degree of circuit selection and guessing entropy evolve between various points in time for these two network load cases. Our analysis focuses on guards rather than on exit nodes: when exit nodes are scarce, no Waterfilling strategy can be applied (or, in other words, the water level reaches the consensus weight of the fastest exit node). Figure~\ref{fig:figure1} uses equations from Section \ref{sec:further_wf} recomputing the bandwidth-weights and applying Waterfilling on the top of them using the constraints \ref{eq:11}, \ref{eq:12}, \ref{eq:13} and \ref{eq:14} from Section~\ref{sec:waterfilling-weights}. For Waterfilling simulations on Figure~\ref{fig:figure2}, we use equations from Section~\ref{sec:waterfilling-weights} because, for this network load case, a balance between the three positions is achieved. For this network scenario, we still have too much bandwidth in the guard position. We apply Waterfilling constraints \ref{eq:11}, \ref{eq:12}, \ref{eq:13} and \ref{eq:14} to obtain a per-node $Wgg_i$. We also apply Waterfilling on the D set since $(E+D) \geq T/3$ as depicted in the appendix.

In all cases, we observe that these metrics favor the Waterfilling scheme over Tor's ABWRS: we have an improvement of around 150 nodes for network case 3aE=SG>M and about 130 nodes for network case 3bE=S. It is an increase of about $\approx 25\%$ for the guessing entropy and an increase of about $\approx 2\%$ for degree of uniformity (which is harder to interpret).  The difference that appears in the metrics between those two network cases can be explained. For 3bE=S, the network achieves a balance for the three positions but not for the network case 3aE=SG>M. It results that the network should achieve better performance for case 3bE=S than case 3aE=SG>M but less diversity because, for 3aE=SG>M we can put in practice the idea introduced in Section \ref{sec:further_wf}. 

We also observe an overall correlation between these two metrics: both are impacted by the uniformity of the probability distribution. But, divergences also happen, as the guessing entropy takes also into account the number of guards and exits: the more we have guards and exits in the network, the higher would be the guessing entropy (except if the distribution is highly non uniform), while such a change may not impact the degree of uniformity. As a result, a variation of nodes in the network between two given moments like the loss of a bunch of guards may result in a drop of the guessing entropy while the degree of uniformity could stay identical.

Entropy measures are a good estimation of anonymity at a given moment for overall usage but they might also be suitable to compare particular user behaviour. Indeed, the guessing entropy would have been smaller in Figures \ref{fig:figure1} and \ref{fig:figure2} if we had considered bittorent download as user model instead of simple HTTP requests, due to the smaller number of exit nodes that allow bittorent ports. On the other hand, the degree of uniformity could fail to show that bittorent users are less protected against end-to-end traffic correlation, simply because the probability distribution of exit nodes allowing bittorent might have the same shape as the probability distribution of exit nodes allowing HTTP requests.

We saw from~\cite{Syverson_why} and~\cite{ccs2013-usersrouted} that
measuring anonymity at a static point in time might not be sufficient:
an adversary may keep monitoring the network for some time and,
depending on the way users update their circuits, have a lower or
higher probability of winning and end-to-end correlation attack at
\emph{some} point in time. But, as Waterfilling does not impact the circuit
evolution strategy, the time evolution is the same for both
approaches.

\begin{figure}[!t]
	\centering
	\includegraphics[scale=0.41]{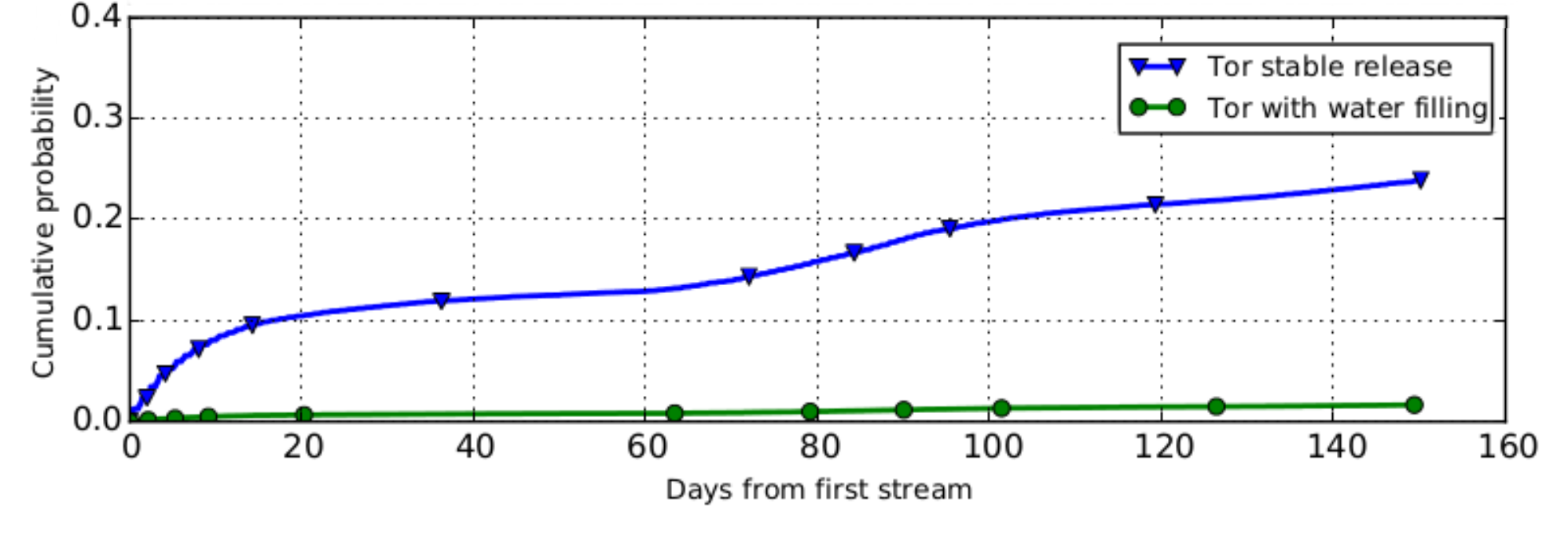}
	\caption{Empirical evaluation of time to first compromise path - With numEntryGuards=3 - We add 1 adversary guard with 480,310 cons weight, 1 adversary exit with 282,607 cons weight}
	\label{comp_tortorwf}
\end{figure}

In the spirit of the work of Johnson \textit{et
al.}~\cite{ccs2013-usersrouted}, we may still wonder about the concrete
impact of a specific adversary over time.
Figure~\ref{comp_tortorwf} shows the time until first compromise
circuit obtained from TorPS with a relay adversary owning a guard with
a consensus weight value of $480,310$ and an exit with consensus weight
value of $282,607$. Both values have been chosen by computing the
consensus weight average of the top-1 guard and exit among the 5-month
time period. In the current design of relay selection, top guards are
a threat to anonymity since they handle the major part of the traffic
in the entry position. Our Waterfilling scheme largely mitigates this
issue: Figure~\ref{comp_tortorwf} shows a drop from $\approx 24 \%$
probability to use a compromised path after 5 months to $\approx 2
\%$. This results from the fact that, with Waterfilling, most of
the capacity of the adversary guard is used for the middle position,
and no single node runs a very significant part of the guard traffic.

\begin{figure}[!t]
	\centering
	\includegraphics[scale=0.41]{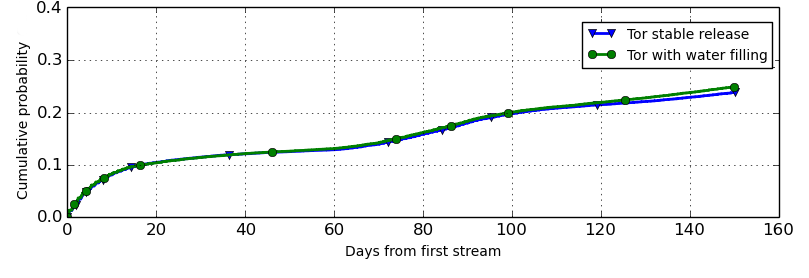}
	\caption{Empirical evaluation of time to first compromise path - 35 adversarial guards cumulating 298,752 cons weight (Waterfilling) versus 1 guard of 480,310 cons weight (Tor ABWRS).}
	\label{comp_tortorwf_35nodes}
\end{figure}

\paragraph*{Effect of Waterfilling on guard rotation}
\label{sec:effect-waterf-guard}
We may also consider the potential interaction that Waterfilling applied on guards could have on guard rotation. In theory, the guard rotation period is randomized between 2 and 3 months. However, in practice, clients rotate guards more often on average due to guard nodes losing their flag or vanishing from the network. Using Waterfilling, we may consider an eventual impact on the average guard rotation period. Indeed, using more smaller guards may affect the expected amount that clients must choose additional guards due to a difference in reachability/uptime with respect to bigger guard nodes. We used TorPS to investigate this phenomenon. Figure~\ref{comp_tortorwf_35nodes} shows a comparison between two simulations. We compare results for classical Tor with 1 adversarial guard of $480,310$ consensus weight with the Waterfilling experiment given $35$ adversarial guards cumulating $298,752$ consensus weight ($62.2\%$ of $480,310$). We use $62.2 \%$ of the consensus weight because at average, guards were used at $62.2 \% $of their bandwidth in the entry position (and $37.8\%$ in the middle position). While with Waterfilling, smaller guards below the water level are used at $100\%$ of their bandwidth in the entry position.
If we were in an ideal situation, where guard nodes never disappears, both curves would perfectly match. Using historical data of the Tor network, Figure~\ref{comp_tortorwf_35nodes} shows that using Waterfilling, the average guard rotation is slightly faster over time, leading to a higher probability of compromise ($+\approx 1\%$). It means that smaller guards tend to be slightly less stable than top guards, but hopefully not that much.

\paragraph*{Security against network adversaries}

\begin{figure}[!t]
	\centering
	\includegraphics[scale=0.40]{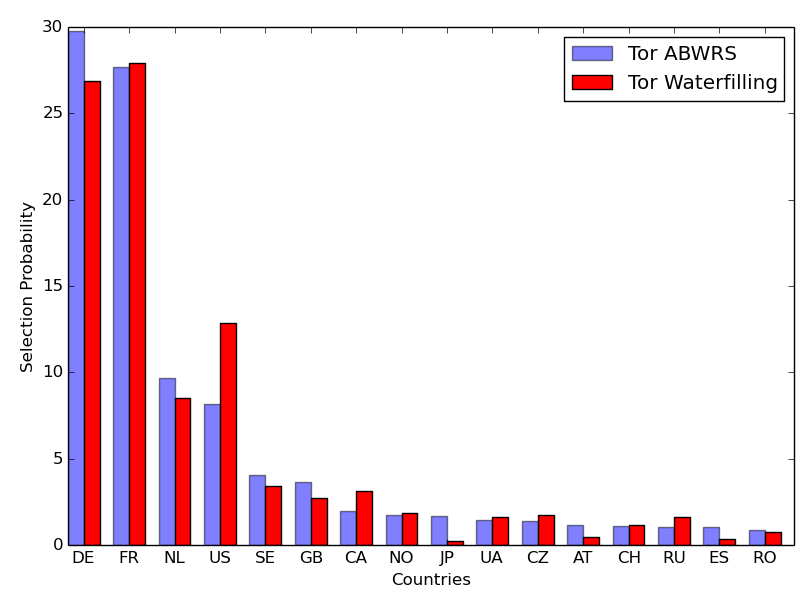}
	\caption{Probability to end up using a guard from the top-16 countries. Like in Section \ref{sec:waterfilling}, the 10th consensus from 25th May 2015 is used.}
	\label{fig:country_diversity}
\end{figure}

Despite the fact that network adversaries are not considered by our
security model, we may still be curious regarding the impact that
Waterfilling has against them.  We use equations \ref{eq:6} from
Section \ref{sec:background} to compute the selection probabilities of
guard nodes. We group them by country and autonomous system, which
allows us to compare the network diversity provided by Waterfilling
and Tor ABWRS against these two types of network adversary. Figure
\ref{fig:country_diversity} shows an overall similarity between the
two schemes. An overall diversity improvement or degradation compared
to Tor ABWRS would be a sign of correlation between the size of relays
and their localization. Here, we see in Figure
\ref{fig:country_diversity} some changes to the few top countries
where the probability to end up in the 2-top countries is a little bit
decreased, from 57.41\% to 54.01\%. However, US jumps from 8.14\% to
13.64\%. It means that among the few top countries, the US hosts the
highest number of guards with a lot of them below the water
level. Table \ref{tab:as_cons_weights} shows {\color{blue}{AS16276}}
(OVH Hosting located in US) jumping from 13.26\% to 18.21\%, which
explains the US jump on Figure \ref{fig:country_diversity}. In the
meantime, {\color{red}{AS12867}} (Online S.A.S located in France)
being the most interesting AS to tap in (containing top guards) is
loosing interest. In any case, it seems hard to draw any clear
conclusions from these numbers, as they are going both ways, and could
change with the evolution of the Tor relay localization and bandwidth.
This is a consequence of our choice to focus the application of
Waterfilling at the relay level rather than at a higher aggregation
level (country, AS, \dots) (see also discussion in
Section~\ref{sec:advers-posit-cost}).

\begin{table}
\begin{tabular}{R{3.5cm}|L{3.5cm}}

 \textit{ABWRS }& \textit{Waterfilling }\\
\hline  \color{red}{AS12876}:\textbf{15.73\%} &  \color{blue}AS16276:\textbf{18.21\%} \\
\hline  \color{blue} AS16276:\textbf{13.26\% }& \color{red}{AS12876}:\textbf{11.24\%} \\
\hline  AS24940:\textbf{10.87\%} & AS24940:\textbf{10.82\%} \\
\hline  AS24961:\textbf{4.42\%}& AS24961:\textbf{3.54\%} \\
\hline  AS8972: \textbf{3.71\%} & AS200130:\textbf{2.08\%} \\
\end{tabular}
\caption{Probability to end up using a guard from the Top-5 AS}
\label{tab:as_cons_weights}
\end{table}

\section{Performance Analysis}
\label{sec:performance}
We now evaluate the performance impact of Waterfilling on a simulated Tor network run in Shadow \cite{shadow-ndss12}.
\subsection{Methodology}
To evaluate the performance, we implemented Waterfilling in the Tor source code and ran several simulations in a virtual network. We constructed a topology scaled down to 796 relays from a consensus in May 2016\footnote{2016-05-28-01-00-00 precisely} (195 guard-flagged nodes, 501 unflagged nodes, 56 exit-flagged nodes, 43 guardexit-flagged nodes, 3 directory authorities and 1 bandwidth authority).
The virtual network topology is built to mimic the current internet structure with geographic clustering by country as explained in Jansen and Hopper \cite{shadow-ndss12}. We generate two classes of experiments. The first one shows simulations in a low network load where we simulated 1125 web clients with 3\% of them performing only bulk transfer. Moreover, we have 75 perf clients for each of 50KiB, 1MiB and 5MiB file size. The overall throughput of this setup is on average $\approx$120 MiB/s, which is proportionally small compared to the traffic load handled by the current Tor network. The second class of experiments shows simulations in a heavy loaded network proportionally close to the throughput of the real Tor network if we don't take into account internal traffic. We run 3000 web clients with 10\% performing bulk transfer. Moreover, we still have 75 perf clients for each of the file size mentioned above. Altogether, those clients push $\approx$ 550 MiB/s, which is $\approx 63\%$ of the total capacity dedicated to the exit position (E+D) in our virtual network.

In both class of experiments, the geographical location of the clients are set up according to Tor's directly connecting user statistics~\cite{tormetrics_clients}. These clients are performing HTTP GET requests to 130 servers with geographic locations assigned using the Alexa Top Sites data set. Therefore, the Tor circuits built by those clients and connecting to the servers should be representative of the real Tor circuits built over the Internet.

\subsection{Analysis}

\begin{figure*}[!t]
	\centering
	\begin{subfigure}[t]{0.32\textwidth}
	 	\centering
	 	\includegraphics[height=1.55in]{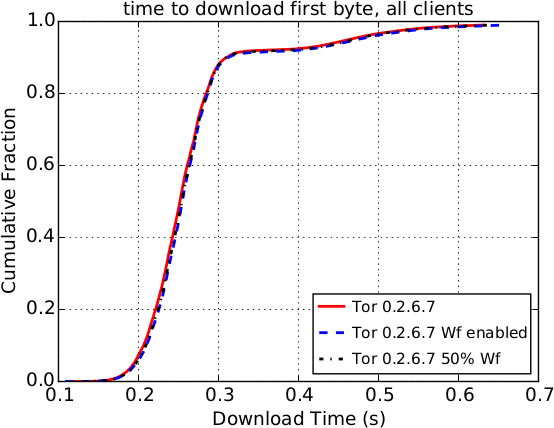}
	 	\caption{Time to first byte (ttfb)}
	 	\label{fig:perf_underload_a}
	\end{subfigure}
	\begin{subfigure}[t]{0.32\textwidth}
	 	\centering
	 	\includegraphics[height=1.55in]{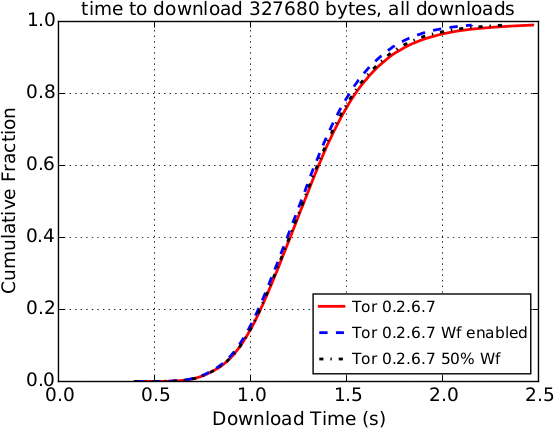}
	 	\caption{Time to download 320 KiB (ttlb)}
	 	\label{fig:perf_underload_b}
	\end{subfigure}
	\begin{subfigure}[t]{0.32\textwidth}
	 	\centering
	 	\includegraphics[height=1.55in]{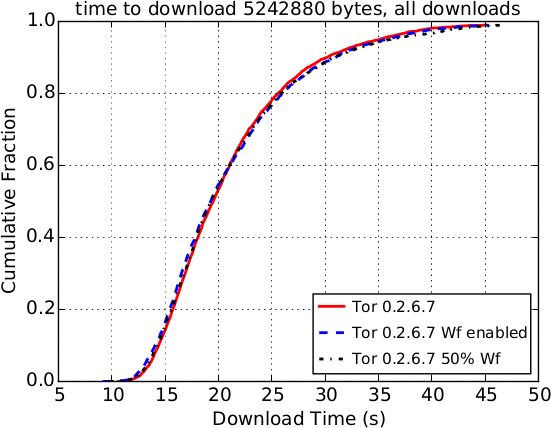}
	 	\caption{Time to download 5 MiB (ttlb)}
	 	\label{fig:perf_underload_c}
	\end{subfigure}
	\caption{Network performance under light load (796 relays, 1125 web clients with 30 bulk clients, 75 perf clients for each of 50KiB, 1MiB and 5MiB file size performing an average throughput of $\approx$ 120 MiB/s altogether). Comparison of Waterfilling fully used on all clients versus Tor's ABWRS used on all clients versus mix 50\% of clients doing Waterfilling and 50\% of clients doing Tor's ABWRS -- The topology generated is scaled down from the first consensus from May 28, 2016. Graphs are plotted until percentile 0.99.}
	\label{fig:perf_underload_1}
\end{figure*}

We study the impact of Waterfilling with two performance metrics: the time to first byte (ttfb) and the time to last byte (ttlb). The first metric shows responsiveness of the network while the second one captures the overall performance. The first class of experiments is under light load, Figure \ref{fig:perf_underload_1} shows no performance discrepancy when using Waterfilling. We may explain latency results (Figure \ref{fig:perf_underload_a}) in the following way: regarding responsiveness of the network, we have two phenomena to consider. relay-based latency and link-based latency. Relay-based latency has shown to be uncorrelated with relay bandwidth in a previous work \cite{congestion-tor12}, thus it does not interfere. Link-based latency is more interesting. To explain why we observe no difference, we may use the following example. Let's suppose that top-bandwidth guards are connected to better links compared to low-bandwidth guards. Therefore, at average, the part of the Tor circuits surrounding guards would suffer from higher latency when using Waterfilling because the clients use guards more uniformly. However, the reverse is applied with middle nodes because we use them less uniformly than current Tor's ABWRS. In average, the part of the Tor circuit surrounding middle nodes would have less latency due to a higher utilization of good links surrounding top nodes. So, with these hypotheses, we should expect more latency through guards and less latency through middles. Because ttfb only cares about the global latency of Tor circuits, these differences tie summed up on average, and we obtain the same result. 

The ttlb metrics (Figures \ref{fig:perf_underload_b}, \ref{fig:perf_underload_c}) confirms our intuition that Waterfilling does not modify the overall performance of the network because it keeps the same total amount of bandwidth for each position, as classical bandwidth-weights. Moreover, we also ran a simulation with half of the clients using Waterfilling weights while the other half are using classical bandwidth-weights. As expected, both techniques can be used simultaneously in the network without impacting the performance. This result allows a smooth transition from Tor ABWRS to Waterfilling or a coexistence if it does not reduce the anonymity set. More details are discussed in Section \ref{sec:deployment}.

\begin{figure*}[!t]
	\centering
	\begin{subfigure}[t]{0.32\textwidth}
	 	\centering
	 	\includegraphics[height=1.55in]{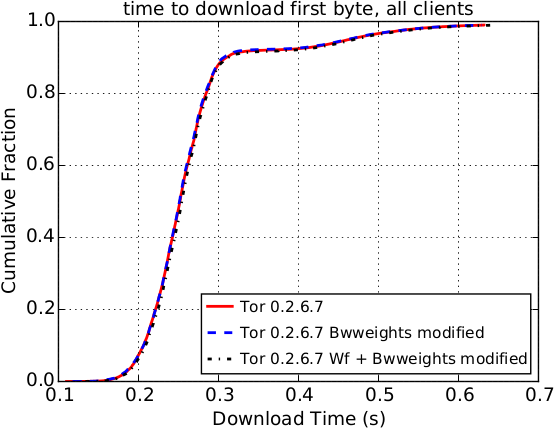}
	 	\caption{Time to first byte}
	\end{subfigure}
	\begin{subfigure}[t]{0.32\textwidth}
	 	\centering
	 	\includegraphics[height=1.55in]{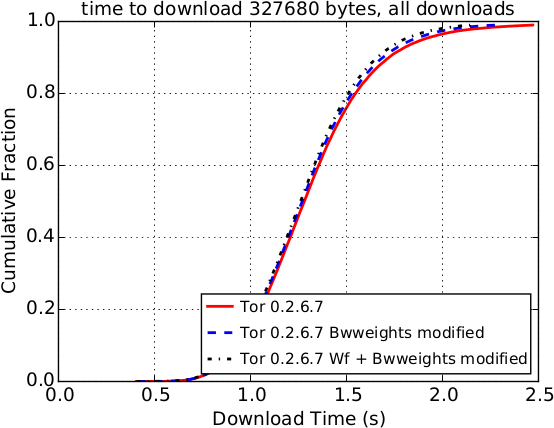}
	 	\caption{Time to download 320 KiB}
	\end{subfigure}
	\begin{subfigure}[t]{0.32\textwidth}
	 	\centering
	 	\includegraphics[height=1.55in]{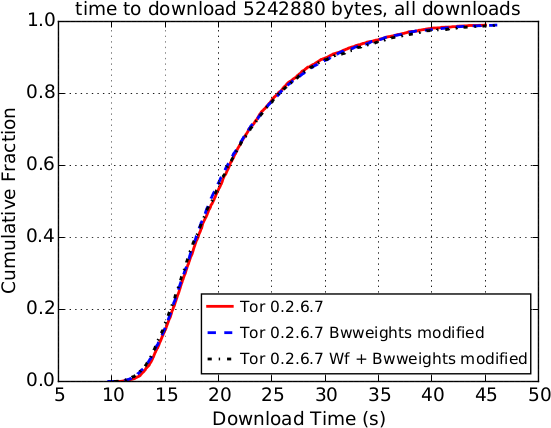}
	 	\caption{Time to download 5 MiB}
	\end{subfigure}
	\caption{Going further with Waterfilling: network performance under light load. Tor with bwweights modified means that we have used equation \ref{eq19} on Tor's ABWRS to obtain the same capacity between entry en exit while pushing what remains to the middle position. Applying Waterfilling on the top of this weight computation gives a more uniform selection probability compared to Waterfilling on the top of classical bandwidth weights. Graphs are plotted until percentile 0.99.}
	\label{fig:perf_underload_2}
\end{figure*}

Section \ref{sec:further_wf} discussed that the classical bandwidth weights can be calculated in a way that improves Waterfilling even more, achieving what we wanted: balancing the Tor network with maximum diversity. The new bandwidth-weights induce the same scarcity between end positions and push what remains to the middle position. Hence, instead of having only one position with low capacity (exit), we now have two positions with low but equal capacity (entry and exit).  In Figure \ref{fig:perf_underload_2}, we investigate if such strategy lowers performance. Under light load, the result show that such strategy does not raise the latency or lower the overall performance of the network. The \textit{Bwweights modified} curves are the bandwidth-weights recalculated using the strategy described in Section \ref{sec:further_wf}.

\begin{figure*}[!t]
	\centering
	\begin{subfigure}[t]{0.32\textwidth}
	 	\centering 
	 	\includegraphics[height=1.55in]{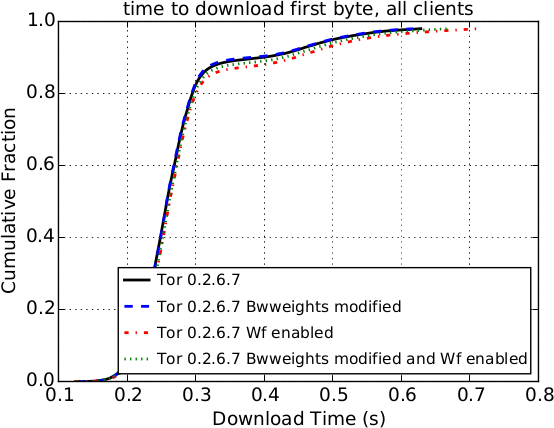}
	 	\caption{Time to first byte}
	\end{subfigure}
	\begin{subfigure}[t]{0.32\textwidth}
	 	\centering
	 	\includegraphics[height=1.55in]{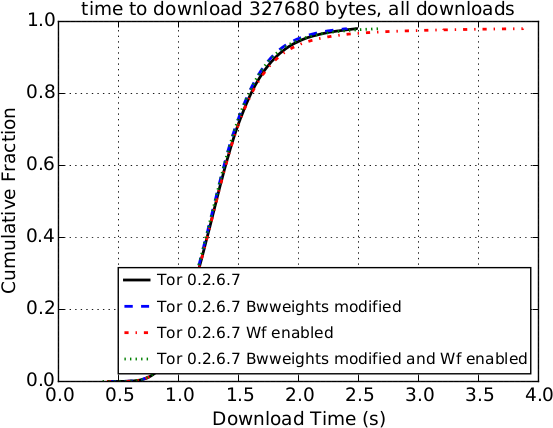}
	 	\caption{Time to download 320 KiB}
	\end{subfigure}
	\begin{subfigure}[t]{0.32\textwidth}
	 	\centering
	 	\includegraphics[height=1.55in]{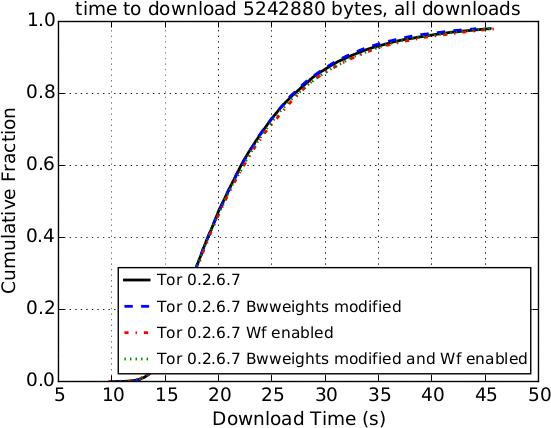}
	 	\caption{Time to download 5 MiB}
	 	\label{fig:perf_loaded_c}
	\end{subfigure}
	\caption{Network performance under loaded traffic ($\approx$550 MiB/s, 63\% of E+D advertised bandwidth). Comparison of Waterfilling fully used on all clients versus Tor's ABWRS fully used on all clients. Graphs are plotted until percentile 0.98. }
	\label{fig:perf_loaded}
\end{figure*}

The second class of results are experiments under a loaded network. Figure \ref{fig:perf_loaded} shows that congestion is increased when using Waterfilling. About 3\% of the worst circuits become a little bit worse, meaning that a few  circuits are more likely to suffer from congestion. However, Figure~\ref{fig:perf_loaded_c} shows that this degradation seems not to appear for large bulk transfers. A surprising and welcoming result comes from the strategy where we apply Waterfilling on the top of recalculated bandwidth-weights. We expected to have more congestion with two positions scarce instead of one. However, the results show that congestion is reduced compared to Waterfilling on the top of classical bandwidth-weights. It seems that pushing as much bandwidth as we can to the middle position has a good impact for the network performance. Since applying Waterfilling on the top this choice improves also anonymity (Section \ref{sec:security}), this result is very welcome. Moreover, pushing as most as we can to the middle position is probably the right thing to do because the balance performed by the bandwidth-weights assumes 3-hop Tor circuits even thought it is not always true. Indeed, Tor circuits carrying hidden service traffic are composed of 2 guards and 4 middles. Moreover, there are also some weird users configurations (such as more than 1 middle node) in the wild. Therefore, we may expect that pushing every exceeding resource to the middle position would give even better performance results on the real Tor network compared to our simulations.

\section{Deployment}
\label{sec:deployment}


Deploying Waterfilling would not raise performance issues and allows a smooth transition between the current path selection and Waterfilling. In theory, the equations from
Sections~\ref{sec:path_selection_details} and
\ref{sec:waterfilling-weights} explain this claim because they still hold in the same way. However, in practice, some other issues might be captured. We conducted the performance analysis using Shadow in Section \ref{sec:performance} and showed classical bandwidth-weights and Waterfilling can be used together by different Tor clients in the
network without disturbing it. The only thing that could be observed
from the relay operator's viewpoint is either a slow transition of their
relay bandwidth consumption from entry to middle or exit to middle if
their relay capacity dedicated to this position is higher than the
water level or a slow transition from middle to entry or middle to
exit if their dedicated relay capacity is smaller than the water
level. The time needed for the transition would depend on Tor users
updating their Tor client. Hopefully, since Tor 5.x.x, the Tor browser
auto-updates, which would bring up to date with Waterfilling a major
fraction of Tor users.

\paragraph*{The deployment security risk}

As explained by Backes \textit{et al.} \cite{backes2015your}, the transition phase or a co-existence of path selection algorithms might be a threat to the anonymity. A bunch of compromised middle nodes might distinguish Waterfilling users from ABWRS users. Moreover, they can also have an idea of what type of service they are accessing, looking at the exit policies. In the case of a transition phase, the first few users switching to the new path selection algorithm are vulnerable. In the case of a co-existence of path selection algorithms, the users are not vulnerable if the transition is finished and both anonymity sets are considered large enough. Hopefully, we could manage to achieve a secure transition phase with a consensus parameter that would be turned on to authorize Waterfilling when enough users have updated.

We specify here a proposal of which type of change should be made in the directory protocol in order for it to be deployed. We suggest modifying the network status document~\cite{dirspec} and add a parameter in \textit{params} called \guillemotleft UseWaterfilling \guillemotright with possible values 0 or 1. When 0 is set, classical bandwidth-weights are used. When 1 is set, the network status document has to give per-node bandwidth-weight values with at most 7 new weights depending on which set(s) Waterfilling is applied to (guards, exits, guards+exits are 3 disjoint sets leading to a possible modification of Wgg, Wmg, Wee, Wme, Wgd, Wed, Wmd). We suggest adding an item called "wfbw" to each router status entry followed by the optional weights in the form $W_{fp}=<value>$.

\section{Limitations and Future Work}
\label{sec:limitations}

The previous sections discussed the impact of Waterfilling on the
design of end-to-end correlation attacks and its expected impact on
the efficiency of the Tor network. We now discuss some open questions
and possible adversarial reactions to Waterfilling.

\subsection{Adversary position and cost structure}
\label{sec:advers-posit-cost}

Our analysis focused on an adversary running an end-to-end correlation
attack by controlling network relay, with a primary focus on the
number of relays that would need to be controlled in order to mount a
successful attack. This captures an adversary who would mount his
attack by hacking into existing relays, assuming that the cost is
proportional to the number of machines to compromise, or paying an
infrastructure for running relays, in which the number of relays to
run would be an important part of the cost (matching the cost
structure of many cloud service providers, for instance). 

Other adversary positions and cost structure could be considered, though. 

\paragraph*{From one to many relays threat}


The analysis in Section~\ref{sec:security} shows that the Waterfilling strategy is very
effective against adversaries targeting top relays: it forces an
adversary to monitor a number of relays in order to achieve what he
could obtain from a single relay right now. This situation holds if we consider relays as individual threats. This is a reasonable assumption in the current path selection since dividing the bandwidth into multiple relays increases the cost without increasing the utility in term of attack effectiveness. 

With Waterfilling, the situation changes. Adversaries could react by taking control (or creating) a large number of relays with a more limited amount of
bandwidth. Assuming that the adversary focuses on relays that offer an
amount of bandwidth close to the ``water level'', we can evaluate
the number of relays that need to be controlled in order to achieve
the same effect as controlling the top guard in the current network.

As explained in Section~\ref{sec:security}, we observed that the top guard achieved an average
consensus weight of 480,310 during the first half of 2015. We also
observe that the average value of $Wgg$ during the same time period is
$62.2\%$, meaning that $480,310*0.622 \approx 298,752$ of the consensus
weight of the top guard is devoted to its guard role. During that same
period, the water level was as low as $8710$. So, in order to achieve
the same consensus weight in total, we see that $298,752/8710 \approx
35$ nodes are needed.

We suspect (and leave for further research) that moving from an
individual threat model to a multiple-relays threat model is
beneficial to the Tor network: running such a strategy would be more
difficult for the adversary:
\begin{itemize}
	\item Regarding budget adversaries, the attacker would have to launch and run his botnet of guard nodes in such a way that would not raise suspicion. The proper way would be to imitate the appearance distribution of other guard nodes, in time and localization. The attacker would need a higher amount of time to be fully operational and would face the difficulty to find proper datacenters to put his nodes. We suspect that Waterfilling increases the effort to set up an effective botnet, hence increasing the budget.
	\item Regarding Intruder adversaries, we see that the adversary now needs to hack into 35 computers instead of one, which must be quite more complex. Using Waterfilling saves the top-bandwidth relays from also being top targets.
\end{itemize}

A potential downside of Waterfilling could appear if we focus on the bandwidth required to mount an attack rather than on the number of nodes that need to be controlled. Indeed, thanks to the unique value $Wgg$ in the current Tor network, only $62.2\%$ of the bandwidth controlled by the attacker is used for the guard role, and this is the useful part for running an end-to-end correlation attack. Waterfilling provides additional utility in terms of attack effectiveness if the attacker splits the bandwidth in multiple relays at the water level.
When using Waterfilling, an adversary aiming for nodes below or up to the water level can hope to have 100\% of its bandwidth allocated to its guard role. As a result, even if the adversary needs to control 35 nodes with Waterfilling instead of one, he only needs to provide $62.2\%$ of the bandwidth that he would need to offer in the current Tor network. An open question is to evaluate the cost of running many nodes to benefit from this additional utility versus the cost of the same utility in the current Tor network.

An interesting adversary reaction against Waterfilling is to take advantage of a botnet structure: among many compromised computers, a few of them might have the required bandwidth and be stable enough to acquire the Guard flag. Such adversary would obtain the additional utility provided by Waterfilling at an additional cost of none, compared to the current path selection. An open question would be to evaluate the number of such compatible botnets in the wild and what would be the cost of running one of them.

One mitigation of this many-relays threat would be to make a new use of the \textit{NodeFamily}: we could apply Waterfilling by considering nodes with the same position and within the same family as a single node. As a result, $n$ guards belonging to the same family and all offering a bandwidth equal to the water level would be selected to offer a \emph{total} guard bandwidth equal to the water level (and not $n$ times that level). 

\paragraph*{Network adversaries}

Analysis in the paper did not show any significant impact of Waterfilling against AS adversaries, given recent localizations of the Tor network relays. But this could change, for the better or for the worse, as new relays appear or disappear. If network adversaries are decided to be the major threat compared to relay-level adversaries, then an interesting open question would be to explore the impact of applying Waterfilling at the AS or country level rather than at the relay level. One contribution to the security analysis in this field \cite{ccs2013-usersrouted, juen2015defending, barton2016denasa} would be to use our guessing entropy to calculate the number of ASes that an adversary must expect to control or compromise before being able to deanonymize a specific circuit. Network adversary reactions should be taken into account in the analysis. Regarding the performance aspect, we conjecture that its marginal impact would be the same if we preserve the total amount of bandwidth per position. Nevertheless, a proper analysis should be conducted.

\subsection{Interaction with other path selection algorithms}
\label{sec:inter-with-other}

%


Waterfilling, like the current path selection algorithm, is used to perform preemptive built of circuits (circuits are built before the user needs them). Some other path selection algorithms have been designed to choose the path between built circuits that maximizes an objective function \cite{congestion-tor12, DBLP:conf/pet/SherrBL09, barton2016denasa}.
An interesting research direction is to look for a combination of different approaches. In this spirit, a combination of destination naive AS-awareness \cite{barton2016denasa} and Waterfilling on relays might lead to better security against relay adversaries and network adversaries with minimal loss of performance. We may even compare this strategy against a mix of Waterfilling on relays and on ASes.

\section{Conclusion}
\label{sec:conclusion}

In this paper, we present a solution to both balance the network and optimize the diversity in endpoints of Tor circuits. This method is called Waterfilling and could be applied in most of the network load cases detailed in the directory server specifications. 

We carry out a security analysis of our scheme with the help of information theoretic metrics and with an empirical estimation over time of the probability until a first path compromise. This estimation is obtained from a modified version of TorPS that implements Waterfilling. We suggest using guessing entropy as a new metric to indicate the strength of the Tor network against relay adversaries at a static point of time. Indeed, guessing entropy captures more aspects than previously used theoretic-information metrics and its meaning may be more informative about the hardness of breaking anonymity, compared to previous notions, like diversity. We show that the guessing entropy increases by about 130 to 150 nodes, or about $25\%$ for any moment throughout the period we consider; meaning that, on average, an adversary would have to corrupt around 130 more nodes before being able to complete a traffic correlation attack at a given time.  

Our security analysis shows also that, when using Waterfilling, a relay adversary needs to control 35 guards in order to obtain the same attack success probability as an adversary controlling the top guard in the current Tor network. We conjecture that taking control of 35 nodes while remaining undetected is a considerably higher challenge than controlling a single one. 
 
We also perform a performance analysis of Waterfilling. Using Shadow, we set up a private network of $\approx$~800 nodes. Our results show a small degradation of performance that might be considered negligible. Indeed, about $97\%$ of Tor circuits constructed with Waterfilling gives the same performance as the circuits constructed by Tor ABWRS. 


\section*{Acknowledgement}
We would like to thanks the anonymous reviewers and our shepherd Nicholas Hopper for their valuable feedbacks and guidance. This work was partially supported by the Innoviris/C-Cure project.

\bibliographystyle{abbrv}
\bibliography{main.bib}

\appendix

\section{Security Models and Metrics}
\subsection{Guessing entropy - example}
\label{appendix:guessing}

Suppose we have a Tor network with 3 guard nodes and 2 exit nodes. From the selection process, suppose we have the following matrix with $p_{i,j}$ the probability to selection guard $i$ with exit $j$

$$ P = \begin{pmatrix}
	1/6 & 1/18\\
	5/18 & 1/3\\
	1/24 & 1/8
\end{pmatrix}$$
Let $G = \emptyset$ the set of prioritized guards and $E=\emptyset$ the set of prioritized exits. $q$ is computed in the following way:

\begin{itemize}	
	\item $q_1 = 0$
	\item $q_2 = max(P) = 1/3, \{g_2\} \cup G, \{e_2\}\cup E$
	\item $q_3 =$ max\_marg\_prob(P)\\
		  \hphantom{$q_3$} $= max(Pr_{G=x}, Pr_{E=y})$\\
		  \hphantom{$q_3$} $= max[max(\underset{y \in E}{\sum} Pr(G=x,$ $E=y) \forall x \notin G),$\\
		  \hphantom{$q_3$ $= max[$}$max(\underset{x \in G}{\sum}Pr(G=x, E=y) \forall y \notin E)]$\\
		  \hphantom{$q_3$} $ =max(1/8, 5/18) = 5/18, \{e_1\}\cup E$
	\item $q_4 =$ max\_marg\_prob(P)\\
		  \hphantom{$q_4$} $ = max(1/6+1/18, \emptyset) = 2/9, \{g_1\} \cup G$\\
	\item $q_5 = 1/24 + 1/8, \{g_3\} \cup G$
\end{itemize}
Then:
$$\sum_{i=1}^{N+K}i.q_i = 3.22$$
\section{Waterfilling Bandwidth-Weights}

\subsection{Waterfilling on the guard+exit flagged nodes}
Waterfilling can also be applied on Guard+Exit flagged nodes when a part of the bandwidth of this pool has to be given to the middle one. In the time period we took to evaluate Waterfilling, it happened when the network load case was 3bE=S.

Suppose there are $K$ nodes with the guard+exit flag, and let $BW_i$ be the
bandwidth of the $i$-th guard+exit flagged node with the highest bandwidth. Then we had the following constraint on the top of classical bandwidth-weights:
\begin{align}
Wd_i*BW_i = Wd_{i+1}*BW_{i+1}~\forall i \in (0, N)
\end{align}
\begin{align}
0 \leq Wd_i \leq 1~\forall i \in (0, K)\\
Wd_i = 1 ~\forall i \in (N+1, K) \\
\sum_{i=0}^{K} Wd_iBW_i = (Wgd + Wed)*D
\end{align}

Solving with these constraints gives $NWd_i$, then we can compute:
$$Wmd_i = 1-Wd_i$$
$$Wgd_i = Wd_i * \frac{Wgd}{Wgd+Wed}$$
$$Wed_i = Wd_i *\frac{Wed }{Wgd+Wed}$$

\end{document}